\documentclass[a4paper, amsfonts, amssymb, amsmath, reprint, showkeys, nofootinbib, twoside]{revtex4-1}
\usepackage[english]{babel}
\usepackage[utf8]{inputenc}
\usepackage[colorinlistoftodos, color=green!40, prependcaption]{todonotes}

\usepackage{amsthm}
\usepackage{mathtools}
\usepackage{physics}
\usepackage{xcolor}
\usepackage{graphicx}
\usepackage[left=23mm,right=13mm,top=35mm,columnsep=15pt]{geometry} 
\usepackage{adjustbox}
\usepackage{placeins}
\usepackage[T1]{fontenc}
\usepackage{lipsum}
\usepackage{csquotes}

\usepackage[pdftex, pdftitle={Article}, pdfauthor={Author}]{hyperref} 
\begin{document}
\title{Lorentz-violation-induced arrival time delay of astroparticles in Finsler spacetime}\thanks{Phys.Rev.D 105 (2022) 124069, \url{https://doi.org/10.1103/PhysRevD.105.124069}}

\author{Jie Zhu}
    \affiliation{School of Physics,
    Peking University, Beijing 100871, China}
    
\author{Bo-Qiang Ma}
    \email[Correspondence email address: ]{mabq@pku.edu.cn}
    \affiliation{School of Physics,
    Peking University, Beijing 100871, China\\
    Center for High Energy Physics, Peking University, Beijing 100871, China\\
    Collaborative Innovation Center of Quantum Matter, Beijing, China}


\begin{abstract}
Finsler geometry is a natural and fundamental generalization of Riemann geometry. The Finsler structure depends on both coordinates and velocities. We present the arrival time delay of astroparticles subject to Lorentz violation
in the framework of Finsler geometry, and the result corresponds to that derived by Jacob and Piran in the standard model of cosmology.
\end{abstract}

\keywords{Lorentz invariance violation, Finsler geometry}

\maketitle

\section{Introduction} \label{sec:introduction}
Lorentz invariance is one of the foundations of the standard model of particle physics. However, in quantum gravity, Lorentz invariance violation (LIV) may happen, and a common feature of many LIV studies is the introduction of modified dispersion relations (MDRs) for elementary particles~\cite{HeMa}. 
When the energies of particles are far below Plank scale ($E_\mathrm{p l}=\sqrt{\hbar c^{5} / G} \sim 1.2 \times 10^{28} \mathrm{eV}$), the modified dispersion relations can be expressed in a general form as the leading term of Taylor series in natural units as
\begin{equation}
    E^2=m^2+p^2\left[1-s_n(\frac{p}{E_{\mathrm{LV},n}})^n \right]=m^2+p^2+\alpha p^{n+2},\label{eq:MDR}
\end{equation}
where $s_n=\pm1$, $E_{\mathrm{LV},n}$ represents the $n$th-order Lorentz violation scale, and $\alpha= -s_n/E_{\mathrm{LV},n}^n$.
The modified dispersion relations bring arrival time differences of astroparticles with different energies. 
Longo~\cite{Longo} and Stodolsky~\cite{Stodolsky} made the earlier proposal to use the time flights of light and neutrinos from the supernova SN1987A to check possible speed anomaly of light and neutrinos. 
Amelino-Camelia {\it et al.}~\cite{method1, method2}  suggested testing Lorentz violation by comparing the arrival times between high- and low-energy photons from gamma-ray bursts, and later many works tested LIV from high-energy photons~\cite{Ellis, shaolijing, zhangshu, xu1, xu2, jie} and neutrinos~\cite{Jacob:2006gn, Amelino-Camelia:2015nqa, Amelino-Camelia:2016fuh, Amelino-Camelia:2016ohi, Huang1, Li}. 
The most widely used formula of the arrival time delay between massless particles with high and low energy~(which is assumed negligible) is derived by Jacob and Piran~\cite{Jacob}, in the standard model of cosmology,
\begin{equation}
    \Delta t=\frac{1+n}{2 H_{0}}\left(\frac{E_\mathrm{obs}}{ E_{\mathrm{LV,}n}}\right)^{n} \int_{0}^{z} \frac{\left(1+z^{\prime}\right)^{n} \mathrm{~d} z^{\prime}}{\sqrt{\Omega_{m}\left(1+z^{\prime}\right)^{3}+\Omega_{\Lambda}}},\label{eq:Jacob}
\end{equation}
where $z$ is the redshift of the source of the two particles, $E_\mathrm{obs}$ is the observed energy of the high-energy particle from Earth equipment, $\Omega_{\rm{m}}$ and $\Omega_{\rm{\Lambda}}$ are universe constants, and $H_0$ is the current Hubble parameter. In their work, they assume the Hamiltonian of the massless particle in terms of the comoving momentum as
\begin{equation}
    \mathcal{H}=\frac{p}{a} \sqrt{1-\left(\frac{p}{a  E_{\mathrm{LV},n}}\right)^{n}},
\end{equation}
where $a=1/(1+z)$ is the cosmological expansion factor, and they assume that the standard relation $v=d\mathcal{H}/dp$ holds; thus they get the equation of the comoving path of the massless particle, and finally they obtain the arrival time delay between massless particles with high energy and normally low energy.

Since particles propagate in the expanding Universe with curvature, it is natural to try to calculate the trajectories of particles in the framework of general relativity (GR) or pseudo-Riemann geometry. However, general relativity respects diffeomorphism invariance as well as local Lorentz invariance, which means that we cannot introduce a modified dispersion relation in the framework of GR; naturally, we try to calculate the trajectories of particles in a more general framework of geometry, i.e., the Finsler geometry~\cite{textbook}.

\section{Finsler Geometry} \label{sec:geometry}

Finsler geometry is a natural and fundamental generalization of Riemann geometry. The Finsler structure depends on both coordinates and velocities. It is defined as a mapping function from a tangent bundle of a manifold to $\mathbb{R}_{0,+}$. In the past few years, more and more research has suggested that new physics may be connected with Finsler geometry, and many kinds of Finsler geometry are studied to pursue new physics~\cite{Foster, Silva, Pfeifer, Hohmann, Silva2}. Many physics models with Lorentz violation are connected with Finsler geometry.
The very special relativity~\cite{VSR} was proved to be a kind of Finsler special relativity~\cite{VSRFinsler}.
The doubly special relativity~\cite{DSR} developed by Amelino-Camelia {\em et al.} can also be incorporated into the framework of Finsler geometry~\cite{DSRFinsler}.
The connection between standard-model extension (SME) and Finsler geometry has been studied in recent years~\cite{SME},
and the classical Lagrangians for SME~\cite{Schreck1,Schreck2} pose the base for constructing Finsler structures.
Li and Chang constructed the theory of gravitation in Berwald–Finsler space~\cite{Chang}. Girelli {\em et al.}~\cite{Girelli} proposed a possible relation between MDRs and Finsler geometry to account for the nontrivial structure of Planckian spacetime. 

Instead of defining an inner product structure over the tangent bundle in Riemann geometry, Finsler geometry is based on the so-called Finsler structure, or Finsler norm $F$ with the property $F(x,\lambda y)=\lambda F(x,y)$ for all $\lambda>0$, where $x \in M$ represents position and $y \equiv \frac{d x}{d \tau}$ represents velocity. The Finsler metric is given as 
\begin{equation}
g_{\mu \nu} \equiv \frac{\partial}{\partial y^{\mu}} \frac{\partial}{\partial y^{\nu}}\left(\frac{1}{2} F^{2}\right).\label{eq:gmn}
\end{equation}
Finsler geometry has its genesis in integrals of the form 
\begin{equation}
    \int_{a}^{b} F\left(x^{1}, \cdots, x^{n} ; \frac{d x^{1}}{d \tau}, \cdots, \frac{d x^{n}}{d \tau}\right) d \tau. \label{eq:F}
\end{equation}
The Finsler structure represents the length element of Finsler space. If $F^2$ is quadratic in $y$, the Finsler metric $g_{\mu \nu}$ is independent of $y$, the Finsler geometry is actually Riemann geometry, and the Finsler metric is said to be Riemann.
To describe the ``1 + 3'' spacetime, instead of Finsler geometry we turn to pseudo-Finsler geometry.
A pseudo-Finsler metric is said to be locally Minkowskian if at every point there is a local coordinate system, such that $F = F (y)$ is independent of the position $x$.

In this work we focus on the geodesic equation of Finsler geometry. The geodesic equation for the Finsler manifold is given as~\cite{textbook}
\begin{equation}
    \frac{d^{2} x^{\mu}}{d \tau^{2}}+2 G^{\mu}=0, \label{eq:geodesic}
\end{equation}
where 
\begin{equation}
    G^{\mu}=\frac{1}{4} g^{\mu \nu}\left(\frac{\partial^{2} F^{2}}{\partial x^{\lambda} \partial y^{\nu}} y^{\lambda}-\frac{\partial F^{2}}{\partial x^{\nu}}\right) \label{eq:spray}
\end{equation}
is called the geodesic spray coefficient. Obviously if $F$ is a Riemann metric, then 
\begin{equation}
    G^{\mu}=\frac{1}{2} \gamma_{\nu \lambda}^{\mu} y^{\nu} y^{\lambda},\label{eq:riemann}
\end{equation}
where $\gamma_{\nu \lambda}^{\mu}$ is the Riemann Christoffel symbol. We can also see that if $F$ is locally Minkowskian, then $G^\mu=0$, and the geodesic equation~(\ref{eq:geodesic}) is actually $\frac{d^{2} x^{\mu}}{d \tau^{2}}=0$.

\section{Pseudo-Finsler Structure of Particles Subject to Lorentz Violation} \label{sec:structure}
A particle moving in a pseudo-Finsler spacetime is described by the action
\begin{equation}
    I=m \int_{a}^{b} F(x, \dot{x}) d \tau. \label{eq:action}
\end{equation}
For a particle with Lorentz violating modified dispersion relation
\begin{equation}
    E^2=m^2+p^2(1+\alpha p^n),\label{eq:dispersion} 
\end{equation}
where $\alpha$ is a parameter with mass dimension $-n$, or $[\alpha]=-n$, 
we derive the pseudo-Finsler geometry of the particle following Ref.~\cite{Girelli}. 

For simplification, we process the procedure in a $``1+1"$ spacetime. As  discussed in Ref.~\cite{Girelli}, we need to introduce a Lagrange multiplier $\lambda$, and letting $p_0=E, p_1=p$, we write the action of the particle as
\begin{equation}
    I=\int\left(\dot{x}^{\mu} p_{\mu}-\lambda\left(p_0^2-p_1^2-\alpha p_1^{n+2}-m^{2}\right)\right) d \tau. \label{eq:l1}
\end{equation}
Defining $\dot{x}^{\mu}=y^\mu$, then we get
\begin{equation}
    I=\int\left({y^0}p_0+y^1p_1-\lambda\left(p_0^2-p_1^2-\alpha p_1^{n+2}-m^{2}\right)\right) d \tau.\label{eq:l2}
\end{equation}
Using Hamilton’s equation, we have
\begin{subequations}
\label{eq:y0y1}
\begin{equation}
y^0=2 \lambda p_0,
\end{equation}
\begin{equation}
y^1=- \lambda (2 p_1+(n+2)\alpha p_1^{n+1}),
\end{equation}
\end{subequations}
and we can solve $p_\mu$ at leading order in $\alpha$ as
\begin{subequations}
\label{eq:p0p1}
\begin{equation}
p_0=\frac{y^0}{2\lambda},
\end{equation}
\begin{equation}
p_1=-\frac{y^1}{2\lambda}+(-1)^n \alpha \frac{(n+2){(y^1)}^{n+1}}{2^{n+2}\lambda^{n+1}}.
\end{equation}
\end{subequations}

We should notice that $p$ in Eq.~(\ref{eq:dispersion}) is the absolute value of the momentum of the particle, so $p_1$ in Eqs.~(\ref{eq:l1}) and~(\ref{eq:l2}) should be its absolute value. For simplification we assume $p_1>0$ in the derivation. Since $p_0>0$ and $\lambda>0$, we can see $y^0>0$ and $y^1<0$ in the derivation. Using Eqs.~(\ref{eq:l2}) and~(\ref{eq:p0p1}), we get the Lagrangian as
\begin{equation}
    L=\frac{(y^0)^2-(y^1)^2}{4\lambda}+\lambda m^2+\alpha \frac{(-y^1)^{n+2}}{2^{n+2}\lambda^{n+1}}+O(\alpha^2).\label{eq:Laglam}
\end{equation}
Varying $\lambda$ in the above Lagrangian, we solve $\lambda$ at leading order in $\alpha$ as 
\begin{equation}
    \lambda=\frac{\sqrt{(y^0)^2-(y^1)^2}}{2 m}+\alpha\frac{(n+1) m^{n-1}(-y^1)^{n+2}}{4\left(\sqrt{(y^0)^2-(y^1)^2}\right)^{n+1}}.\label{eq:lambda}
\end{equation}
Using the relation we obtain the particle Lagrangian at leading order in $\alpha$ as 
\begin{equation}
    L=m\sqrt{(y^0)^2-(y^1)^2}+\alpha m^{n+1}\frac{(-y^1)^{n+2}}{2\left(\sqrt{(y^0)^2-(y^1)^2}\right)^{n+1}}, \label{eq:Lag}
\end{equation}
and the pseudo-Finsler norm
\begin{equation}
    F=\sqrt{(y^0)^2-(y^1)^2}+\alpha m^{n}\frac{(-y^1)^{n+2}}{2\left(\sqrt{(y^0)^2-(y^1)^2}\right)^{n+1}}. \label{eq:Finsler}
\end{equation}
As mentioned above, we should remind the reader that we present the derivation under the assumption of $p_1>0$ and $y^1<0$. We can also process the same procedure under the assumption of $p_1<0$ and $y^1>0$, which means that Eq.~(\ref{eq:dispersion}) becomes
\begin{equation}
    E^2=m^2+p^2(1+\alpha (-p)^n),
\end{equation}
and finally we get the full form of the pseudo-Finsler norm as
\begin{equation}
    F=\sqrt{(y^0)^2-(y^1)^2}+\alpha m^{n}\frac{\left|y^1\right|^{n+2}}{2\left(\sqrt{(y^0)^2-(y^1)^2}\right)^{n+1}}.
\end{equation}
We can write the pseudo-Finsler norm in $1 + 3$ spacetime as
\begin{equation}
    F=\sqrt{\eta_{\mu\nu}y^{\mu}y^{\nu}}+\alpha m^{n}\frac{(y^a y^a)^{\frac{n+2}{2}}}{2\left(\eta_{\mu\nu}y^{\mu}y^{\nu}\right)^\frac{n+1}{2}},\label{eq:finsler4d}
\end{equation}
where $\eta_{\mu\nu}=\mathrm{diag}(1,-1,-1,-1)$, and $a$ is a spatial index, which is summed over. This result is compatible with the result obtained by the SME community~\cite{SME} in a different way. 

Just as discussed in Ref.~\cite{Girelli}, even assuming a universal coefficient $\alpha$ in Eq.~(\ref{eq:dispersion}), still the MDR corresponds to a pseudo-Finsler norm that is mass-dependent; this means that particles with different masses see different pseudo-Finsler structures. That is because pseudo-Finsler norms have no scale embedded in them as a consequence of $F(x,\lambda y)=\lambda F(x,y)$. If we introduce a dimensional $\alpha$ in a locally Minkowskian pseudo-Finsler norm, there must be another dimensional constant to cancel the scale, and that is the mass of the particle since there is no position coordinate appearing in the pseudo-Finsler norm. It seems that we cannot calculate the trajectories of massless particles with MDRs in pseudo-Finsler geometry; however, we can deal with massless particles just as we do in Riemann geometry, and we will discuss this in the next section.

It is natural to assume that a particle moves along a geodesic in pseudo-Finsler spacetime. As we can see from Eq.~(\ref{eq:finsler4d}), the pseudo-Finsler norm of the particle is independent of the position coordinates $x^\mu$. As discussed in Sec.~\ref{sec:geometry}, the geodesic equation is just $\frac{d^{2} x^{\mu}}{d \tau^{2}}=0$, which means that a free particle in a locally Minkowskian pseudo-Finsler spacetime propagates with a constant speed. From Eq.~(\ref{eq:y0y1}), we can get the speed of the particle
\begin{equation}
    v=\left|\frac{\dot{x}^1}{\dot{x}^{0}}\right|=\left|\frac{y^1}{y^{0}}\right|=\frac{p}{E}\left[ 1+\alpha \frac{n+2}{2} p^{n}\right],
\end{equation}
which is related to the second of Eq.~(2) in Ref.~\cite{Schreck1} and the same as derived from the assumption $v=\partial E/\partial p$ in conventional studies.

\section{Time Delay in Expanding Universe} \label{sec:time}

Now we turn to the expanding Universe of the standard model of cosmology. To calculate the motion of a particle, we need to obtain the pseudo-Finsler structure and solve the geodesic equation corresponding to the pseudo-Finsler structure. Before we get into this procedure, we simply look back at how we solve the particle propagation problem in the Riemann spacetime.

The expanding Universe can be described by the Friedmann-Robertson–Walker~(FRW) metric, and in a $1 + 1$ Riemann spacetime the length element is $ds=\sqrt{dt^2-a(t)^2dx^2}$, or $F_R=\sqrt{(y^0)^2-(a(x^0)y^1)^2}$ in a Finsler way, where $a(t)$ is the cosmological expansion factor, $x^0=t$, $x^1=x$, $y^0=dx^0/d\tau$, and $y^1=dx^1/d\tau$. Let us set present time as $t=0$, thus we have $a(0)=1$. For $a(t)$, the Hubble parameter $H$, and the redshift $z$, there are relations that $a=\frac{1}{1+z}$, $H=\frac{a'(t)}{a(t)}$, and $dz=-(1+z)Hdt$. 

Assume that a particle starts to move at $t=-T$ and $x=X$ with redshift $z_0$ and reaches us at $t=0$ and $x=0$, and we can measure its energy and momentum $E_\mathrm{obs}$ and $P_\mathrm{obs}$. Obviously, we have $y^0=dt/d\tau>0$, $y^1=dx/d\tau<0$, and $dx/dt<0$. The geodesic equations of the FRW metric are shown as
\begin{subequations}
\label{eq:eqFRW}
\begin{equation}
\Ddot{x}^0+a(x^0)a'(x^0)(\dot{x}^1)^2=0,\label{eq:FRW1}
\end{equation}
\begin{equation}
\Ddot{x}^1+2\frac{a'(x^0)}{a(x^0)}\dot{x}^1\dot{x}^0=0.\label{eq:FRW2}
\end{equation}
\end{subequations}
From Eq.~(\ref{eq:FRW2}), we can get
\begin{equation}
    y^1=\dot{x}^1=\frac{C_1}{a(t)^2},\label{eq:FRWsol1}
\end{equation}
and combining Eq.~(\ref{eq:FRWsol1}) and Eq.~(\ref{eq:FRW1}), we can get
\begin{equation}
    y^0=\dot{x}^0=\sqrt{\epsilon+\frac{C_1^2}{a(t)^2}},\label{eq:FRWsol2}
\end{equation}
where $C_1$ and $\epsilon$ are integration constants with $C_1<0$. As we know, if the particle is massless, such as a photon, then $\epsilon=0$, and for a massive particle, if $\tau$ is set to be the proper time, then $\epsilon=1$. Instead of using the common sense above, here we determine the constants with boundary conditions. At $t=0$, $a(t)=1$, the velocity of the particle is $v=\left|y^1/y^0\right|=\frac{-C_1}{\sqrt{\epsilon+C_1^2}}$, and thus
\begin{equation}
    P_\mathrm{obs}=\frac{m v}{\sqrt{1-v^2}}=\frac{-m C_1}{\sqrt{\epsilon}}.
\end{equation}
So we can let
\begin{equation}
    \epsilon=\frac{C_1^2 m^2}{P_\mathrm{obs}^2},
\end{equation}
and combing Eqs.~(\ref{eq:FRWsol1}) and~(\ref{eq:FRWsol2}), we have
\begin{equation}
    \frac{dx}{dt}=\frac{y^1}{y^0}=-\frac{P_\mathrm{obs}}{a(t)\sqrt{m^2 a(t)^2+P_\mathrm{obs}^2}}.
\end{equation}
If $P_\mathrm{obs}\gg m$, the above equation becomes
\begin{equation}
    \frac{dx}{dt}=-\frac{1}{a(t)},
\end{equation}
and the equation above is exactly the same as the equation for massless particles.

Now we turn to the Finsler expanding universe.
In the Finsler expanding universe, the Minkowski metric is replaced by the pseudo-Riemann metric of the FRW spacetime. 
As we can see, the FRW metric can be derived from replacing $(dx^\alpha)^2$ with $a(t)^2 (dx^\alpha)^2$ in the Riemann Minkowski metric, where $\alpha$ is a space index, or replacing $y^\alpha$ with $a(x^0) y^\alpha$. It is natural to think in this way because $a(t)$ describes how the space expands and it should be multiplied to every space component in the metric. Thus we can write the pseudo-Finsler norm from Eq.~(\ref{eq:finsler4d}) as
\begin{equation}
    F'=\sqrt{g_{\mu\nu}y^{\mu}y^{\nu}}+\alpha m^{n}\frac{a(x^0)^{n+2}(y^a y^a)^{\frac{n+2}{2}}}{2\left(g_{\mu\nu}y^{\mu}y^{\nu}\right)^\frac{n+1}{2}},\label{eq:finsler4dFRW}
\end{equation}
where $g_{\mu\nu}=\mathrm{diag}(1,-a(x^0)^2,-a(x^0)^2,-a(x^0)^2)$. Considering a particle propagating in a $1 + 1$ spacetime, and assuming the motion of the particle described as above, we can get the pseudo-Finsler norm for the particle as
\begin{equation}
    F'=\sqrt{(y^0)^2-(a(x^0)y^1)^2}+\alpha m^{n}\frac{(-a(x^0)y^1)^{n+2}}{2\left(\sqrt{(y^0)^2-(a(x^0)y^1)^2}\right)^{n+1}}, \label{eq:finslerFRW}
\end{equation}
and the factor $(-1)^{n+2}$ appeals in Eq.~(\ref{eq:finslerFRW}) because $y^1<0$. Now we can get the geodesic equation for the particle at leading order
in $\alpha$ as
\begin{widetext}
\begin{subequations}
\label{eq:eqFRWFinsler}
\begin{equation}
\dot{y}^0+a(x^0)a'(x^0)(y^1)^2 + 
\alpha m^n  \frac{(n+2)a'(x^0) a(x^0)^{n+1}(-y^1)^{n+2}\left[(n-1)(y^0)^2+a(x^0)^2(y^1)^2\right]}{2\left[ (y^0)^2-a(x^0)^2(y^1)^2 \right]^{\frac{n+2}{2}}}=0,\label{eq:FRWFinsler1}
\end{equation}
\begin{equation}
\dot{y}^1+2\frac{a'(x^0)}{a(x^0)}y^0y^1 - 
\alpha m^n \frac{n(n+2)a'(x^0)a(x^0)^{n-1}(y^0)^3(-y^1)^{n+1}}{2\left[ (y^0)^2-a(x^0)^2(y^1)^2 \right]^{\frac{n+2}{2}}}=0.\label{eq:FRWFinsler2}
\end{equation}
\end{subequations}
\end{widetext}

Equation~(\ref{eq:eqFRWFinsler}) is much more complicated than Eq.~(\ref{eq:eqFRW}). However, we get its symbolic solution at leading order in $\alpha$. To solve the geodesic equation, we assume that the solution has the form
\begin{subequations}
\label{eq:form}
\begin{equation}
    y^1=\frac{C_1}{a(x^0)^2}+\alpha m^n f(\tau),\label{eq:form1}
\end{equation}
\begin{equation}
    y^0=\sqrt{\epsilon+\frac{C_1^2}{a(x^0)^2}}+\alpha m^n g(\tau), \label{eq:form2}
\end{equation}
\end{subequations}
where $C_1<0$. Combing Eqs.~(\ref{eq:eqFRWFinsler}) and~(\ref{eq:form}), and expanding the equation to $O(\alpha^2)$, we can get the equations for $f(\tau)$ and $g(\tau)$.
We should notice that $f'(\tau)=\frac{df}{da}\frac{da}{dx^0}\frac{dx^0}{d\tau}=a'(x^0)y^0\frac{df}{da}$ and the same for $g(\tau)$. Using this, we can get the equation for $f(a)$ and $g(a)$ as
\begin{widetext}
\begin{subequations}
\label{eq:fg}
\begin{equation}
    f'(a)+\frac{2}{a}f(a)-\frac{n(n+2)(-C_1)^{n+1}(\epsilon a^2+C_1^2)}{2 a^{n+5} \epsilon^{\frac{n+2}{2}}}=0,\label{eq:f}
\end{equation}
\begin{equation}
    \sqrt{\epsilon+\frac{C_1^2}{a^2}}g'(a)-\frac{C_1^2}{a^2\sqrt{\epsilon a^2 +C_1^2}}g(a)+
    \frac{2C_1}{a}f(a)+\frac{(n+2)(-C_1)^{n+2}((n-1)\epsilon a^2+n C_1^2)}{2a^{n+5} \epsilon^{\frac{n+2}{2}}}=0,\label{eq:g}
\end{equation}
\end{subequations}
\end{widetext}
and the solution for Eq.~(\ref{eq:fg}) is
\begin{widetext}
\begin{subequations}
\label{eq:solfg}
\begin{equation}
    f(a)=\frac{C_2}{a^2}-\frac{(-C_1)^{n+1}\left((n+2)\epsilon a^2+n C_1^2\right)}{2 \epsilon^{\frac{n+2}{2}}a^{n+4}},\label{eq:solf}
\end{equation}
\begin{equation}
    g(a)=\frac{C_3 a}{\sqrt{\epsilon a^2+C_1^2}}
    +\frac{C_1C_2}{a\sqrt{\epsilon a^2+C_1^2}}
    +\frac{(-C_1)^{n+2}\left((n+1)\epsilon a^2+n C_1^2\right)}{2\epsilon^\frac{n+2}{2}a^{n+3}\sqrt{\epsilon a^2+C_1^2}},\label{eq:solg}
\end{equation}
\end{subequations}
\end{widetext}
where $C_2$ and $C_3$ are integration constants. We will see that when the energy of the particle in much bigger than its mass, $C_2$ and $C_3$ do not contribute to observables, so that we can set $C_2=C_3=0$, but now we still keep it. Finally we get the solution of Eq.~(\ref{eq:eqFRWFinsler}) at leading order in $\alpha$ as
\begin{widetext}
\begin{subequations}
\label{eq:soly0y1}
\begin{equation}
    y^1=\frac{C_1}{a^2}+\alpha m^n\left[\frac{C_2}{a^2}-\frac{(-C_1)^{n+1}\left((n+2)\epsilon a^2+n C_1^2\right)}{2 \epsilon^{\frac{n+2}{2}}a^{n+4}}\right],\label{eq:soly1}
\end{equation}
\begin{equation}
    y^0=\sqrt{\epsilon+\frac{C_1^2}{a^2}}+\alpha m^n\left[\frac{C_3 a}{\sqrt{\epsilon a^2+C_1^2}}
    +\frac{C_1C_2}{a\sqrt{\epsilon a^2+C_1^2}}
    +\frac{(-C_1)^{n+2}\left((n+1)\epsilon a^2+n C_1^2\right)}{2\epsilon^\frac{n+2}{2}a^{n+3}\sqrt{\epsilon a^2+C_1^2}}\right].\label{eq:soly0}
\end{equation}
\end{subequations}
\end{widetext}

Here we discuss what is an observable in this pseudo-Finsler spacetime. Obviously, the coordinate $x^\mu$, the energy $E$, and the momentum $p$ are observables. $y^\mu=\frac{dx^{\mu}}{d\tau}$ are not an observable for we can change $\tau$ at will, but the ratio of $y^a/y^0=dx^a/dx^0$ is an observable and actually it represents the speed defined by how we measure it. From Eqs.~(\ref{eq:p0p1}) and ~(\ref{eq:lambda}) we can see the energy and the momentum can also be calculated by $y^a/y^0$, whitch means for the solution Eq.~(\ref{eq:soly0y1}) only the ratio $y^1/y^0=dx/dt$ has physical meaning, and at leading order in $\alpha$ the ratio is
\begin{widetext}
\begin{equation}
    \frac{dx}{dt}=
    \frac{C_1}{a\sqrt{\epsilon a^2+C_1^2}}+\alpha m^n \left[
    \frac{\epsilon C_2 a}{(\epsilon a^2+C_1^2)^\frac{3}{2}}
    -\frac{C_1C_3 a}{(\epsilon a^2+C_1^2)^\frac{3}{2}}
    -\frac{(n+2)(-C_1)^{n+1}\epsilon^\frac{2-n}{2}}{2 a^{n-1} (\epsilon a^2+C_1^2)^\frac{3}{2}}
    -\frac{(n+1)(-C_1)^{n+3}\epsilon^{-\frac{n}{2}}}{2 a^{n+1}(\epsilon a^2+C_1^2)^\frac{3}{2}}
    \right].\label{eq:speed}
\end{equation}
\end{widetext}
Dimensional analysis on Eq.~(\ref{eq:speed}) shows that $[\epsilon]=2[C_1]$. Just like how we deal with the geodesic equation of the FRW metric, let
\begin{equation}
    \epsilon=\frac{C_1^2m^2}{P_o^2},\label{eq:ep}
\end{equation}
where $[P_o]=[m]$, then we get
\begin{widetext}
\begin{equation}
    \frac{dx}{dt}=-\frac{P_o}{a\sqrt{m^2a^2+P_o^2}}+\alpha\left[
    -\frac{C_2 m^{n+2}P_oa}{C_1(m^2a^2+P_o^2)^\frac{3}{2}}
    +\frac{C_3 m^n P_o^3 a}{C_1^2 (m^2a^2+P_o^2 )^\frac{3}{2}}
    -\frac{(n+2) m^2 P_o^{n+1}}{2 a^{n-1} (m^2a^2+P_o^2 )^\frac{3}{2}}
    -\frac{(n+1)P_o^{n+3}}{2a^{n+1}(m^2a^2+P_o^2 )^\frac{3}{2}}
    \right],\label{eq:speedPm}
\end{equation}
\end{widetext}
and we will see soon that $P_o$ is actually the observed momentum $P_\mathrm{obs}$ of the particle.
We can see that when $P_o\gg m$, the first three terms in the square brackets are suppressed in comparison to
the fourth term in the square brackets in Eq.~(\ref{eq:speedPm}), and we finally get 
\begin{equation}
    \frac{dx}{dt}=-\left(\frac{1}{a}+\frac{n+1}{2}\alpha P_o^n\frac{1}{a^{n+1}} \right).\label{eq:finalspeed}
\end{equation}

Equation~(\ref{eq:finalspeed}) is quite simple, and we also find that $C_2$ and $C_3$ disappear in the equation, which means that these two constants have no contribution to the observables, so we can set $C_2=C_3=0$. Considering the boundary condition at $t=0$, we have $a=1$, $p_0=E_\mathrm{obs}$ and $p_1=P_\mathrm{obs}$. Combining Eqs.~(\ref{eq:soly0y1}),~(\ref{eq:ep}),~(\ref{eq:p0p1}) and~(\ref{eq:lambda}), we have
\begin{subequations}
\begin{equation}
    E_\mathrm{obs}={\sqrt{m^2+P_o^2}+\frac{\alpha P_o^{n+2}}{2\sqrt{m^2+P_o^2}}}+O(\alpha^2),
\end{equation}
\begin{equation}
    P_\mathrm{obs}=P_o+O(\alpha^2),
\end{equation}
\end{subequations}
and we prove the assertion that $P_o$ is actually the observed momentum of the particle at $t=0$. Consider that $P_\mathrm{obs}=E_\mathrm{obs}+O(\alpha)$ and change the variable $t$ to redshift $z$, then Eq.~(\ref{eq:finalspeed}) can be rewritten as
\begin{equation}
    \frac{dx}{dz}=\frac{1}{H(z)}+\frac{(n+1)\alpha E_\mathrm{obs}^n}{2} \frac{(1+z)^n}{H(z)} .\label{eq:finalspeedz}
\end{equation}
Following the work of Jacob and Piran\cite{Jacob}, we get the time delay formula as
\begin{equation}
    \Delta t=\frac{n+1}{2}\alpha E_\mathrm{obs}^n \int_0^z \frac{(1+z)^n}{H(z)}dz,
\end{equation}
using $H(z)=H_0\sqrt{\Omega_{m}\left(1+z\right)^{3}+\Omega_{\Lambda}}$, then
\begin{equation}
    \Delta t=\alpha E_\mathrm{obs}^n \frac{n+1}{2H_0}\int_0^z \frac{(1+z')^n}{\sqrt{\Omega_{m}\left(1+z^{\prime}\right)^{3}+\Omega_{\Lambda}}}dz',
\end{equation}
which is exactly the same as the time delay induced by the Lorentz violation effect between two particles with different energies in the expanding Universe,  i.e., Eq.~(\ref{eq:Jacob}) obtained by Jacob and Piran~\cite{Jacob} in the standard model of cosmology. 
From Eq.~(\ref{eq:finalspeed}), we see that if $\alpha>0$, then high-energy particles propagate faster and arrive earlier, and if $\alpha<0$, high-energy particles propagate slower and arrive later.

\section{Conclusion and Discussion}\label{sec:conclusion}
In this work, we derive the pseudo-Finsler structure of a particle subject to Lorentz violation from a general modified dispersion relation as Eq.~(\ref{eq:MDR}) following the work of Ref.~\cite{Girelli}. We perform a detailed calculation of the trajectory of the particle subject to Lorentz violation in the expanding Universe by the geodesic equation of the pseudo-Finsler structure and calculate the arrival time delay between particles with high energy and normally low energy. Surprisingly, the formula of the arrival time delay induced by the Lorentz violation effect between two particles with different energies is exactly the same as Jacob and Piran~\cite{Jacob} got, in a different way from the standard model of cosmology. The consistency of the results suggests that Finsler geometry is a good effective theory to describe quantum gravity. 
Since Finsler geometry provides a means to describe particle propagation in a non-Riemann spacetime, e.g., when a particle is subject to Lorentz violation or something along these lines,
the method performed in this work may be applied to other questions, such as how gravitational lensing or a black hole can influence 
the propagation of a particle subject to Lorentz violation, and this is exactly what other theories cannot deal with. 
\\

\section*{ACKNOWLEDGMENTS} \label{sec:acknowledgements}
This work is supported by National Natural Science Foundation of China (Grant No.~12075003).


\end{document}